\documentclass[journal]{IEEEtran}
\usepackage{amsmath}
\usepackage{amssymb}
\usepackage{setspace}
\usepackage{epsfig}
\usepackage{epstopdf}
\usepackage{rotating}
\usepackage{color}
\usepackage{colortbl}
\usepackage{graphicx}
\usepackage{subfig}
\usepackage{ifthen}
\usepackage{alltt}
\usepackage{fancyhdr}
\usepackage{verbatim}
\usepackage{moreverb}
\usepackage{cite}
\usepackage{multirow}
\usepackage{bigstrut}
\def\e{\begin{equation}}
\def\f{\end{equation}}
\def\_#1{{\bf #1}}

\def\.{\cdot}

\def\l#1{\label{#1}}
\def\r#1{(\ref{#1})}
\def\=#1{\overline{\overline{#1}}}

\def\aeeb{\={\alpha}_{\rm ee}}
\def\aemb{\={\alpha}_{\rm em}}
\def\ameb{\={\alpha}_{\rm me}}
\def\ammb{\={\alpha}_{\rm mm}}
\def\aeeo{\alpha_{\rm ee}^{\rm co}}
\def\aeer{\alpha_{\rm ee}^{\rm cr}}
\def\aemo{\alpha_{\rm em}^{\rm co}}
\def\aemr{\alpha_{\rm em}^{\rm cr}}
\def\ameo{\alpha_{\rm me}^{\rm co}}
\def\amer{\alpha_{\rm me}^{\rm cr}}
\def\ammo{\alpha_{\rm mm}^{\rm co}}
\def\ammr{\alpha_{\rm mm}^{\rm cr}}

\def\aeeoh{\widehat{\alpha}_{\rm ee}^{\rm co}}
\def\aeerh{\widehat{\alpha}_{\rm ee}^{\rm cr}}
\def\aemoh{\widehat{\alpha}_{\rm em}^{\rm co}}
\def\aemrh{\widehat{\alpha}_{\rm em}^{\rm cr}}
\def\ameoh{\widehat{\alpha}_{\rm me}^{\rm co}}
\def\amerh{\widehat{\alpha}_{\rm me}^{\rm cr}}
\def\ammoh{\widehat{\alpha}_{\rm mm}^{\rm co}}
\def\ammrh{\widehat{\alpha}_{\rm mm}^{\rm cr}}
\def\aeebh{\={\widehat{\alpha}}_{\rm ee}}
\def\aembh{\={\widehat{\alpha}}_{\rm em}}
\def\amebh{\={\widehat{\alpha}}_{\rm me}}
\def\ammbh{\={\widehat{\alpha}}_{\rm mm}}

\def\It{\overline{\overline{I}}_{\rm t}}
\def\Jt{\overline{\overline{J}}_{\rm t}}

\def\=#1{\overline{\overline{#1}}}

\begin{document}

\title{Total absorption of electromagnetic waves in ultimately thin layers}

\author{Y. Ra'di,~\IEEEmembership{Student member,~IEEE}, V.S. Asadchy, S.A. Tretyakov,~\IEEEmembership{Fellow,~IEEE}
\thanks{Y. Ra'di, V.S. Asadchy, S.A. Tretyakov are with the Department of Radio Science and Engineering/SMARAD Center of Excellence, Aalto University, P.~O.~Box~13000, FI-00076 AALTO, Finland. Email: younes.radi@aalto.fi.}
}

\markboth{Ra'di \MakeLowercase{\textit{et al.}}: Total absorption of electromagnetic waves in ultimately thin layers}%
{Ra'di \MakeLowercase{\textit{et al.}}: Total absorption of electromagnetic waves in ultimately thin layers}
\maketitle

\begin{abstract}
We consider single-layer arrays of electrically small lossy
particles that completely absorb electromagnetic waves at normal
incidence. Required conditions for electromagnetic properties of
bi-anisotropic particles have been identified in the most general
case of uniaxial reciprocal and nonreciprocal particles. We consider
the design possibilities offered by the particles of all four
fundamental classes of bi-anisotropic inclusions: reciprocal chiral
and omega particles and nonreciprocal Tellegen and moving particles.
We also study the reflection/transmission properties of asymmetric
structures with different properties when illuminated from the
opposite sides of the sheet. It has been found that it is possible
to realize single-layer grids which exhibit the total absorption
property when illuminated from one side but are totally transparent
when illuminated from  the other side (an ultimately thin isolator).
Other possible properties are co-polarized or twist polarized
reflection from the side opposite to the absorbing one. Finally, we
discuss possible approaches to practical realization of particles
with the properties required for single-layer perfect absorbers and
other proposed devices.

\end{abstract}



\section{Introduction}
\label{sec:introduction}

\IEEEPARstart{I}{n} this paper we study possible  approaches to the
design of electrically thin layers (sheets) which would behave as
{\em perfect absorbers} for normally-incident electromagnetic plane
waves. We say that absorption in a layer at some frequency is
``perfect'' or ``total'', if all incident power is dissipated in the
layer. This implies that both reflection and transmission
coefficients are equal to zero. In this study we will consider only
the case of normal incidence, thus, this term should not be confused
with the {\em perfectly matched layer} or PML, which implies zero
reflection coefficient at any incident angle and for any
polarization of the incident wave.

The theory and design  of absorbers for electromagnetic radiation
has a long history and there exists a large variety of designs,
especially for microwave frequencies (see, e.g. \cite{RCS,
radar_absorbers2, radar_absorbers}). However, in most of these
designs the absorbing structure is backed by a reflecting wall
(usually a metal surface), because most often the goal is to reduce
microwave reflections from metal structures. Recently, there has
been considerable interest in thin absorbing layers for situations
where there is no reflector behind, so that the electromagnetic
properties of the object which one wants to ``hide" can be
arbitrary. Naturally, a thin reflector can be incorporated in the
absorber structure, but often it is desirable to allow off-band
electromagnetic waves to pass through the structure or avoiding
conductors is one of the application requirements. Also, for
infrared and visible-light applications the use of  perfect
reflectors as parts of absorbing layers is not practically possible
except if the use of electrically thick layers of photonic crystals
is allowed in design. Electrically thin matched absorbers  can be
realized in many ways, but, to the best of our knowledge, only a
very limited set of opportunities has been explored so far. One
known possibility is to combine two thin metamaterial layers with
contrasting material parameters \cite{Bilotti2006} or combine a thin
resistive sheet with an array of small resonant split rings (which
realize the necessary magnetic response) \cite{Bilotti2011}.
A fundamental limitation on the bandwidth of metal-backed
absorbers has been discussed in \cite{Rozanov}. A review of recently
introduced multilayer absorbers can be found in \cite{Watts}. Here
we will not consider such two- or multilayer structures,
concentrating on the basic and fundamentally simplest case of a
single sheet with properly designed properties. These single-layer
absorbers provide ultimately thin design solutions, because the
layer thickness cannot be made smaller than just one layer of
particles (molecules).

Conceptually, the simplest possible thin absorbing sheet is a
uniform or composite layer of electrically negligible thickness
(impedance sheet).  In this case, the incident electric field
induces an infinitesimally thin sheet of electric current in the
layer, which eventually leads to dissipation of the incident power,
if the layer is lossy. However, it is obvious that in this case the
absorbed power can reach only one half of the incident power, and
the total absorption is not possible (e.g.,
\cite{add1,Pozar_array}). This follows from the fact that the
induced current sheet symmetrically radiates plane waves in the
forward and back directions. Zero transmission coefficient implies
that the amplitude of this secondary wave behind the sheet equals to
that of the incident wave (so that the two waves cancel each other
behind the sheet), but this means that the reflection coefficient
equals unity in the amplitude. Thus, in order to enable total
absorption, we need to allow also magnetic current to be induced in
the layer. Strictly speaking, this implies that the layer thickness
cannot be negligibly small (electrically), at least if no natural
magnetics are used, but it can be still made very small compared
with the wavelength.

In view of practical  requirements in realization of layers with
desired electromagnetic response, calling for the use of composite
structures, we do not model the layer as a homogeneous sheet
described by some surface susceptibility or impedance, but assume
from the beginning that the layer is a composite structure: a single
layer of small polarizable particles. Engineering these inclusions,
we can tune the reflection and transmission responses of the
composite layer. Such artificial sheets with engineered
electromagnetic properties are called ``metasurfaces'' or
``metasheets'', see recent reviews \cite{Holloway,Shalaev_review}.
The absorber designs which we will develop here will give
the required polarizabilities of individual inclusions together with
the appropriate array period. In order to have full design
freedom in defining how the induced electric and magnetic moments of
the absorbing dipolar particles depend on the incident fields, we
assume that the particles are the most general bi-anisotropic
particles, possibly nonreciprocal.

Here, we will consider only  the case of normal plane-wave
incidence. In practice, performance stability for oblique incidence
is an important issue. Conventional absorbers are made of material
layers of considerable electrical thickness, and this thickness is
the key parameter defining the resonant frequency. Because this
thickness changes when the incidence angle of the incident wave
deviates from the normal, a shift in the frequency of the absorption
maximum is expected. The ultimately thin absorbing layers proposed
in this paper are expected to be more stable for changes of the
incident angle, but a separate study is needed to understand the
angular dependence of response of these new structures.

The condition for total absorption of normally  incident plane waves
by a single infinite periodic array of electric and magnetic dipoles
is known from the antenna theory \cite{Pozar_array}. Let us assume
that an infinite array with the period $a$ ($a$ is smaller than the
wavelength in the surrounding space) in each unit cell contains one
isotropic particle in which the incident electric and magnetic
fields induce electric dipole moment $\_p$ and magnetic moment
$\_m$. The two moments will be orthogonal: electric moment along the
incident electric field and magnetic moment along the magnetic
field. Arrays of both moments will create secondary plane waves, and
in the forward direction the secondary electric field amplitude
reads \e E_{\rm forward}={-j\omega \over 2S}\left(\eta_0 p +{1\over
\eta_0} m\right)\f where $S=a^2$ is the unit-cell area, so that
$j\omega p/S$ is the surface-averaged electric current density and
$j\omega m/S$ is the magnetic current surface density.
$\eta_0=\sqrt{\mu_0/\epsilon_0}$ is the wave impedance of the
surrounding space. Derivation of these formulas for plane-wave
fields created by planar sheets of electric and magnetic currents
can be found e.g. in \cite{Felsen_M}. In the opposite direction (the
reflection direction), the same induced electric and magnetic
currents generate a plane wave with the amplitude \e E_{\rm
back}={-j\omega \over 2S}\left(\eta_0 p -{1\over \eta_0} m\right)\f
Now, we see that it is possible to choose the moments so that the
secondary field would cancel the incident field $E_{\rm inc}$ in the
forward direction (zero transmission coefficient) and at the same
time the secondary field would be zero in the back direction (zero
reflection coefficient). Obviously, the conditions are
\e p={-jS\over \omega \eta_0}E_{\rm inc},\qquad m=\eta_0^2 p \l{ccc}
\f This arrangement of electric and magnetic current sheets is in
fact a Huygens surface, and for volumetric material layers that
would correspond to materials with equal relative permittivity and
permeability. We note in passing that the use of volumetric
materials with matched wave impedance in absorbers is well known,
see e.g. \cite{RCS,Sol} or \cite[ch.~12]{basic}.

Thus,  a simple approach to realization of total absorbing layers is
to arrange electrically and magnetically polarizable particles in a
dense lattice and tune the polarizabilities so that \r{ccc} are
satisfied. However, this is not the only possible approach. We only
need to ensure that the dipole moments have the required
values, but the particles in which these dipole moments are induced
can be any electrically small objects which one  can describe as
dipole scatterers. We expect that there should be considerable
design freedom and possibilities for realizing additional
practically useful properties if we do not restrict the design space
by the simplest case of small electrically and magnetically
polarizable scatterers (like small magnetodielectric spheres, for
example).

In this paper we consider planar layers formed by electrically small
particles modeled by the most general linear relations  between the
induced dipole moments $\_p$ and $\_m$ and the local fields
$\mathbf{E}_{\rm loc}$ and $\mathbf{H}_{\rm loc}$ at the positions
of the particles:
\begin{equation}
\left[ \begin{array}{c} \mathbf{p} \\ \mathbf{m}\end{array} \right]
=\left[ \begin{array}{cc} \aeeb& \aemb\\
\ameb& \ammb
\end{array} \right]\cdot \left[ \begin{array}{c} \mathbf{E}_{\rm
loc} \\ \mathbf{H}_{\rm loc}\end{array} \right].
\label{eq:e1}\end{equation} Although for the desired operation of
the layer the induced dipole moments must satisfy the same
conditions of the Huygens sheet \r{ccc}, the actual design space is
vastly larger, since we can exploit the magneto-electric coupling
parameters $\aemb$ and
$\ameb$ to bring the induced moments to
the desired balance and required amplitudes. Furthermore, additional
functionalities will become possible, as we will see in the
following.

While the simple and well-known solution in form of electric and
magnetic dipole particles corresponds to  simple magnetodielectric
layers with $\epsilon_r=\mu_r$ (if we think about layers of
homogeneous materials), the general case of bi-anisotropic
polarizabilities of individual particles corresponds to
bi-anisotropic absorbing layers. In the past, chiral absorbing
layers were studied in detail
\cite{chiral87,chiral89,chiral89a,chiral92,chiral96,cloete,Koschny},
but only for metal-backed volumetric layers. The use of omega
coupling phenomenon for matched absorber layers was explored in
\cite{basic, omega, reference2}, but also only for material layers
on perfectly conducting surfaces. Recently, different kinds of
absorbers have been proposed to absorb  electromagnetic waves in
microwave or optical spectra
\cite{Sajuyigbe1,Zhou1,Korolev1,Yuan1,Jiang1,Cui1,Shvets1}. As it
was mentioned, most of these absorbers are backed with a metal sheet
which limits their functionality for the wave coming from the other
side. These absorbers contain more than one layer of particles and
they are designed so that to absorb the wave from one side while
they have some uncontrollable properties for the wave coming from
the other side. Here, we answer important questions: How one can
realize single-layer perfect absorbers from one side of the sheet
and what functionalities can be realized for waves coming from the
other side? Of course, this implies that there is no metal (PEC)
ground plane as a part of the absorbing structure.

Here, we study the possible use of single arrays of
bi-anisotropic particles of all known classes: reciprocal chiral
and omega and nonreciprocal Tellegen and ``moving" particles
\cite{classes,basic}. Also in this paper we consider the use of
particles which have hybrid electromagnetic properties of several
classes, e.g. ``moving" chiral and Tellegen omega particles.
The electromagnetic coupling of the artificial omega-Tellegen
particle was measured experimentally in \cite{mTellegen}.  It was
shown that nonreciprocal electromagnetic coupling really exists in
the particle and the electromagnetic coupling coefficient is
commensurate by magnitude with the electric and magnetic
polarizabilities. Described implementation of the particle implies
presence of external magnetic field bias (3570 Oe in
\cite{mTellegen}). The used material of ferrite inclusions was
yttrium iron garnet. 
\section{Total absorption in arrays of general bi-anisotropic particles}
\subsection{Effective polarizability dyadics of particles in periodic arrays}
In this paper, we consider thin absorbers for normally incident
plane waves and concentrate on uniaxial structures, isotropic in the
plane of the layer. This property ensures that the absorber
functions for arbitrary polarized incident plane waves. The
orientation of the absorbing sheet in space is defined by the unit
vector $\_z_0$, orthogonal to its plane.  The layer consists of an
array of electrically small uniaxial particles. As discussed above,
total absorption requires at least electric and magnetic dipole
moments induced in the particles, and the requirement of ultimately
small thickness means that higher-order multipoles are negligible.
Thus, we assume that the particles are bi-anisotropic particles
characterized by four dyadic polarizabilities: electric, magnetic,
electromagnetic, and magnetoelectric, which relate local
electromagnetic fields to the induced electric and magnetic dipole
moments as in (\ref{eq:e1}).

The uniaxial symmetry allows only isotopic response and rotation
around the axis $\_z_0$. Thus, all the polarizabilities in
(\ref{eq:e1}) take the forms:
\begin{equation}
\begin{array}{c}
\aeeb=\aeeo\It+\aeer\Jt,\qquad \displaystyle
\ammb=\ammo\It+\ammr\Jt\\\vspace*{.1cm}\displaystyle
\aemb=\aemo\It+\aemr\Jt,\qquad \displaystyle
\ameb=\ameo\It+\amer\Jt,
\end{array}\label{eq:g1}
\end{equation}
where indices ${\rm co}$ and ${\rm cr}$ refer to the symmetric and antisymmetric
parts of the corresponding dyadics, respectively.
$\It=\overline{\overline{I}}-\mathbf{z}_0\mathbf{z}_0$
is the transverse unit dyadic and
$\Jt=\mathbf{z}_0\times\It$
is the vector-product operator. The particles are arranged in a
square lattice with the unit cell of the size $a\times a$. The grid
is excited by an arbitrary polarized plane wave with the electric
and magnetic fields of $\mathbf{E}_{\rm inc}$ and $\mathbf{H}_{\rm
inc}$, respectively, which are uniform in the array plane (normal
incidence). In this situation, the induced dipole moments are the
same for all particles. We assume that the grid period $a$ is
smaller than the wavelength, so that no grating lobes are generated.

The local fields exciting the particles are the sums of the external
incident field and the interaction field caused by the induced
dipole moments in other particles:
\begin{equation}
\begin{array}{c}
\mathbf{E}_{\rm loc}=\mathbf{E}_{\rm{inc}}+\overline{\overline{\beta}}_{\rm e}\cdot\mathbf{p}
\vspace*{.2cm}\\\displaystyle
\mathbf{H}_{\rm loc}=\mathbf{H}_{\rm inc}+\overline{\overline{\beta}}_{\rm m}\cdot\mathbf{m}  ,
\end{array}\label{eq:h1}
\end{equation}
where  $\overline{\overline{\beta}}_{\rm e}$ and
$\overline{\overline{\beta}}_{\rm m}$ are the interaction constants. These
dyadic coefficients are proportional to the two-dimensional unit
dyadic $\It$. Explicit analytical expressions for the interaction
constants can be found in  \cite{basic}.

Equations (\ref{eq:e1}) and (\ref{eq:h1}) can be re-written as
relations between  the induced dipole moments and the incident
fields:
\begin{equation}
\left[ \begin{array}{c} \mathbf{p} \\ \mathbf{m}\end{array} \right]
=\left[ \begin{array}{cc}
\aeebh &
\aembh\\\amebh&
\ammbh \end{array} \right]\cdot
\left[ \begin{array}{c} \mathbf{E}_{\rm inc} \\ \mathbf{H}_{\rm
inc}\end{array} \right] , \label{eq:j1}
\end{equation}
where the effective polarizabilities (marked by hats) include the
effects of particle interactions in the array. Explicit formulas for
the effective polarizabilities in terms of the individual
polarizabilities and interaction constants are given in
\cite{Teemu}:
\begin{equation}
\hspace*{-.2cm}\begin{array}{c}
\aeebh\!=\!\left(\It\!-\!\aeeb\!\cdot\!\overline{\overline{\beta}}_{\rm e}\!-\!\aemb\!\cdot\!\overline{\overline{\beta}}_{\rm m}\!\cdot\!(\It\!-\!\ammb\!\cdot\!\overline{\overline{\beta}}_{\rm m})^{-1}\!\cdot\!\ameb\!\cdot\!\overline{\overline{\beta}}_{\rm e}\right)^{-1}
\vspace*{.2cm}\\\displaystyle
.\left(\aeeb+\aemb\cdot\overline{\overline{\beta}}_{\rm m}\cdot(\It-\ammb\cdot\overline{\overline{\beta}}_{\rm m})^{-1}\cdot\ameb\right)

\vspace*{.4cm}\\\displaystyle
\aembh\!=\!\left(\It\!-\!\aeeb\!\cdot\!\overline{\overline{\beta}}_{\rm e}\!-\!\aemb\!\cdot\!\overline{\overline{\beta}}_{\rm m}\!\cdot\!(\It\!-\!\ammb\!\cdot\!\overline{\overline{\beta}}_{\rm m})^{-1}\!\cdot\!\ameb\!\cdot\!\overline{\overline{\beta}}_{\rm e}\right)^{-1}
\vspace*{.2cm}\\\displaystyle
.\left(\aemb+\aemb\cdot\overline{\overline{\beta}}_{\rm m}\cdot(\It-\ammb\cdot\overline{\overline{\beta}}_{\rm m})^{-1}\cdot\ammb\right)

\vspace*{.4cm}\\\displaystyle

\amebh\!=\!\left(\It\!-\!\ameb\!\cdot\!\overline{\overline{\beta}}_{\rm e}\!\cdot\!(\It\!-\!\aeeb\!\cdot\!\overline{\overline{\beta}}_{\rm e})^{-1}\!\cdot\!\aemb\!\cdot\!\overline{\overline{\beta}}_{\rm m}\!-\!\ammb\!\cdot\!\overline{\overline{\beta}}_{\rm m}\right)^{-1}
\vspace*{.2cm}\\\displaystyle
.\left(\ameb+\ameb\cdot\overline{\overline{\beta}}_{\rm e}\cdot(\It-\aeeb\cdot\overline{\overline{\beta}}_{\rm e})^{-1}\cdot\aeeb\right)

\vspace*{.4cm}\\\displaystyle

\ammbh\!=\!\left(\It\!-\!\ameb\!\cdot\!\overline{\overline{\beta}}_{\rm e}\!\cdot\!(\It\!-\!\aeeb\!\cdot\!\overline{\overline{\beta}}_{\rm e})^{-1}\!\cdot\!\aemb\!\cdot\!\overline{\overline{\beta}}_{\rm m}\!-\!\ammb\!\cdot\!\overline{\overline{\beta}}_{\rm m}\right)^{-1}
\vspace*{.2cm}\\\displaystyle
.\left(\ammb+\ameb\cdot\overline{\overline{\beta}}_{\rm e}\cdot(\It-\aeeb\cdot\overline{\overline{\beta}}_{\rm e})^{-1}\cdot\aemb\right).
\end{array}\label{eq:l1}
\end{equation}
Because the interaction constants are diagonal dyadics, the symmetry
properties of the effective polarizabilities are the same as for the
individual particle polarizabilities (as defined in (\ref{eq:g1})):
\begin{equation}
\begin{array}{c}
\aeebh=\aeeoh\It+\aeerh\Jt,\qquad \displaystyle
\ammbh=\ammoh\It+\ammrh\Jt\\\vspace*{.1cm}\displaystyle
\aembh=\aemoh\It+\aemrh\Jt,\qquad\displaystyle
\amebh=\ameoh\It+\amerh\Jt\displaystyle.
\end{array}\label{eq:k1}
\end{equation}
This can be checked by substituting (\ref{eq:g1}) in (\ref{eq:l1}).
\subsection{Reflection and transmission coefficients}
In the theory of absorbing sheets, we will distinguish between
illuminations of the sheet from its two opposite sides, along
$-\mathbf{z}_0$ and $\mathbf{z}_0$. In the rest of the paper, we
will use double signs for these two cases, where the top and bottom
signs correspond to the incident plane wave propagating in
$-\mathbf{z}_0$ and $\mathbf{z}_0$ directions, respectively.

In the incident plane wave, the electric and magnetic fields satisfy
\begin{equation}
\mathbf{H}_{\rm
inc}=\mp\frac{1}{\eta_0}\Jt\cdot\mathbf{E}_{\rm
inc}. \label{eq:m1}\end{equation} Thus, the dipole moments in
(\ref{eq:j1}) can be written as
\begin{equation}
\displaystyle \left[ \displaystyle\begin{array}{c} \mathbf{p} \\
\mathbf{m}\end{array} \right] =\left[\displaystyle
\begin{array}{c}\displaystyle
\aeebh\mp\frac{1}{\eta_0}\aembh\cdot(\mathbf{z}_0\times\It)\vspace{.1cm}\vspace*{.2cm}\\\displaystyle
\amebh\mp\frac{1}{\eta_0}\ammbh\cdot(\mathbf{z}_0\times\It)
\end{array}\right]\cdot  \begin{array}{c} \mathbf{E}_{\rm inc}
\end{array}. \label{eq:n1}\end{equation}
Secondary plane waves (reflected and transmitted) are generated by
surface-averaged current densities
\begin{equation}
\displaystyle
\mathbf{J}_{\rm e}=\frac{j\omega}{S}\mathbf{p}, \qquad
\mathbf{J}_{\rm m}=\frac{j\omega}{S}\mathbf{m}.
\label{eq:o1}
\end{equation}
Radiation from infinite sheets of electric and magnetic currents can
be easily solved  \cite{Teemu} from the Maxwell equations:
\begin{equation}
\begin{array}{l}
\displaystyle
\mathbf{E}_{\rm r}=-\frac{j\omega}{2S}\left\{\left[\eta_0\aeeoh\pm \aemrh\pm \amerh-\frac{1}{\eta_0} \ammoh\right]\It\right.\vspace*{.2cm}\\\displaystyle
\hspace*{1.6cm}\left.+\left[\eta_0\aeerh\mp \aemoh\mp \ameoh-\frac{1}{\eta_0} \ammrh\right]\Jt\right\}\cdot\mathbf{E}_{\rm inc}
\end{array}\label{eq:q1}
\end{equation}
\begin{equation}
\begin{array}{l}
\displaystyle
\mathbf{E}_{\rm t}=\left\{\left[1-\frac{j\omega}{2S}\left(\eta_0\aeeoh\pm\aemrh\mp\amerh
+\frac{1}{\eta_0}\ammoh\right)\right]\It\right.\vspace*{.2cm}\\\displaystyle
\hspace*{.3cm}\left.
-\frac{j\omega}{2S} \left[\eta_0\aeerh\mp\aemoh
\pm\ameoh+\frac{1}{\eta_0} \ammrh\right] \Jt\right\}\cdot\mathbf{E}_{\rm inc}.
\end{array}\label{eq:s1}
\end{equation}

Using these general expressions for the reflected and transmitted
fields from general bi-anisotropic planar arrays, we are ready to
study how we can make these fields equal to zero, as required for
perfect absorbers.
\subsection{General conditions for total absorption}

\subsubsection{Total absorption from both sides of the sheet}

The definition of a perfect absorber implies that
 \begin{equation}
\begin{array}{c}
\hspace*{.5cm}\mathbf{E}_{\rm r}=0,\hspace*{.5cm}\mathbf{E}_{\rm t}=0
\end{array}\label{eq:t1}.
\end{equation}
Equating to zero the expressions in square brackets in (\ref{eq:q1})
and (\ref{eq:s1}), we arrive to sufficient conditions for total
absorption of arbitrarily polarized incident plane waves:
\begin{equation}
\begin{array}{c}
\displaystyle
\eta_0\aeeoh\pm \aemrh\pm\amerh-\frac{1}{\eta_0} \ammoh=0
\vspace*{.2cm}\\\displaystyle
\eta_0\aeerh\mp \aemoh\mp \ameoh-\frac{1}{\eta_0} \ammrh=0
\vspace*{.2cm}\\\displaystyle
\eta_0\aeeoh\pm\aemrh\mp\amerh+\frac{1}{\eta_0}\ammoh=\frac{2S}{j\omega}
\vspace*{.2cm}\\\displaystyle
\eta_0\aeerh\mp\aemoh\pm\ameoh+\frac{1}{\eta_0} \ammrh=0.
\end{array}\label{eq:u1}
\end{equation}
Because in the expressions for the reflected and transmitted fields
(\ref{eq:q1}) and (\ref{eq:s1}) the terms proportional to $\It$
and $\Jt$ are orthogonal, these conditions are also the necessary
conditions for total absorption. The exception for the last
statement is the case of circularly polarized incident waves, when
these conditions are sufficient but not necessary, opening still
more design possibilities if only circularly polarized waves should
be absorbed totally. For circularly polarized incidence, the total
absorption conditions read
\begin{equation}
\begin{array}{l}
\displaystyle
\eta_0\aeeoh\pm \aemrh\pm\amerh-\frac{1}{\eta_0} \ammoh
\vspace*{.2cm}\\\displaystyle
\hspace*{1.6cm}=(\pm j)  \left[
\eta_0\aeerh\mp \aemoh\mp \ameoh-\frac{1}{\eta_0} \ammrh\right]
\vspace*{.4cm}\\\displaystyle
 1-\frac{j\omega}{2S}\left(\eta_0\aeeoh\pm\aemrh\mp\amerh
+\frac{1}{\eta_0}\ammoh\right)
\vspace*{.2cm}\\\displaystyle
\hspace*{1.4cm}= (\mp j) \frac{j\omega}{2S} \left[\eta_0\aeerh\mp\aemoh
\pm\ameoh+\frac{1}{\eta_0} \ammrh\right]    .
\end{array}\label{eq:u1CP}
\end{equation}
Here $\pm j$ coefficients correspond to the two orthogonal
polarizations of the incident circularly polarized fields.

In the following, we will use the general sufficient conditions
(\ref{eq:u1}) which ensure total absorption for arbitrary
polarization of the incident waves. As one can see, these conditions
connect symmetric and antisymmetric parts of the electric and
magnetic polarizabilities to the antisymmetric and symmetric parts
of the cross-coupling polarizabilities, respectively. This is a very
important point because for reciprocal particles (for example,
arbitrary shaped metal or dielectric particles) the antisymmetric
parts of the electric and magnetic polarizabilities are zero (e.g.,
\cite{basic}) which limits symmetric components of electromagnetic
coupling dyadics. Furthermore, we see from (\ref{eq:u1}) that for
zero reflection it is not necessary to have absorption inside the
particles, while it is necessary for zero transmission (note the
imaginary quantity in the right-hand side of the third equation).

Let us first analyze layers which exhibit the total absorption
property from both sides of the sheet. In this case, conditions
(\ref{eq:u1}) should hold for both choices of the $\pm$ signs, and
we find that all the magnetoelectric coefficients must vanish: \e
\aemrh=\amerh=\aemoh=
\ameoh=0.\f Thus, we conclude that the only
possible realization of total absorbers in form of a single layer of
particles is the use of electrically and magnetically polarizable
uniaxial particles with the polarizabilities balanced as in a
Huygens' pair: \e \aeeoh={S\over j\omega
\eta_0}, \qquad \ammoh=\eta_0^2
\aeeoh\l{both_sides}\f (and all the other
polarizability components equal zero). The effective
polarizabilities which include the effect of particle interactions
in the array should be, thus, purely imaginary, corresponding to a
resonance where the particles show purely absorptive properties.

The relations between the collective polarizabilities and the
polarizabilities of the same particles in free space (\ref{eq:l1})
in this special case simplify to \e {1\over
\aeeoh}={1\over \aeeo}-\beta_{\rm e},\qquad
{1\over \ammoh}={1\over
\ammo}-\beta_{\rm m} . \f Using the known expression for the
interaction constants in regular dipolar arrays
\cite[eq.~(4.89)]{modeboo}, we can find the required particle
polarizabilities in free space: \e {1\over \aeeo}={\rm
Re}(\beta_{\rm e})+j{k^3\over 6\pi\epsilon_0}+j{\omega \eta_0   \over 2
S},\f \e {1\over \ammo}={\rm Re}(\beta_{\rm m})+j{k^3\over
6\pi\mu_0}+j{\omega \over 2S\eta_0}.\f We again see that the
reactive response of the individual particles should be such that
together with the reactive part of the interaction field a resonance
condition is satisfied. We can also check that the amplitude of the
secondary plane waves created by the two dipolar arrays of the
perfect absorber equal to one half of the incident field amplitude:
\e E_{\rm sc}=-{\eta_0\over 2}{j\omega p\over S}=-{\eta_0\over
2}{j\omega \over S}\aeeoh E_{\rm inc}=-{1\over
2}E_{\rm inc}\f (we have substituted $\aeeoh$
from \r{both_sides}). The field created by the magnetic-dipole array
has the same amplitude. In the forward directions, the sum of these
two plane waves compensates the incident field, and in the
reflection direction these two plane waves are out of phase and the
sum is zero.

\subsubsection{Total absorption from one side of the sheet}

Next, we consider single-layer sheets which work as total absorbers
only from one side and study what functionalities can be engineered
for illumination from the opposite side. From (\ref{eq:u1}), we know
that the presence of cross-coupling polarizabilities (as well as the
anti-symmetric parts of the electric and magnetic polarizabilities)
causes asymmetry in interactions of the sheet with incident waves
coming from the opposite directions. Let us assume that we satisfy
(\ref{eq:u1}) for waves incident from one of the two sides. This
corresponds to conditions

\begin{equation}
\begin{array}{c}
\displaystyle
\eta_0\aeeoh-\frac{1}{\eta_0} \ammoh=\mp ( \aemrh+\amerh)
\vspace*{.2cm}\\\displaystyle
\eta_0\aeerh-\frac{1}{\eta_0} \ammrh=\pm (\aemoh+ \ameoh)
\vspace*{.2cm}\\\displaystyle
\eta_0\aeeoh+\frac{1}{\eta_0}\ammoh= \frac{2S}{j\omega}\mp (\aemrh-\amerh)
\vspace*{.2cm}\\\displaystyle
\eta_0\aeerh+\frac{1}{\eta_0} \ammrh=\pm (\aemoh-\ameoh).
\end{array}\label{eq:teem5}
\end{equation}
Here, as above, the upper sign corresponds to $-\_z_0$ directed and
the lower sign to the oppositely-directed incident plane waves.
Using the same conditions for total absorption for the opposite
incidence direction (taking the lower signs in (\ref{eq:u1}),
(\ref{eq:q1}), and (\ref{eq:s1})), we find the reflected and
transmitted electric fields for the same sheet when the incidence is
from the other side:

\begin{equation}
\begin{array}{l}
\displaystyle
\mathbf{E}_{\rm r}=\frac{j\omega}{S}\left\{\pm \left[ \aemrh+\amerh\right]\It\mp\left[ \aemoh+ \ameoh\right]\Jt\right\}\cdot\mathbf{E}_{\rm inc}
\end{array}
\label{eq:teem6}
\end{equation}
\begin{equation}
\begin{array}{l}
\displaystyle
\mathbf{E}_{\rm t}=\frac{j\omega}{S}\left\{\pm \left[ \aemrh-\amerh\right]\It\mp\left[ \aemoh- \ameoh\right]\Jt\right\}\cdot\mathbf{E}_{\rm inc}.
\end{array}\label{eq:teem7}
\end{equation}
These equations show that tuning the layer to act as a perfect
absorber from one side, it is possible to realize some special
properties (in reflection and transmission) from the other side.

To study these properties, we begin with the case of reciprocal
structures. In this case, the electric and magnetic polarizabilties
are symmetric dyadics ($\aeerh=0$,
$\ammrh=0$) and the fields coupling coefficients
 satisfy \e
\aemoh=-\ameoh,\qquad
\aemrh=\amerh,\f
corresponding to chiral and omega couplings \cite{basic}. Due to the
reciprocity, the transmission coefficient is zero for waves incident
from both sides. The second equation in (\ref{eq:teem5}) is
satisfied identically, and from the last one, we see that the
chirality parameter $\ameoh=0$. Thus, if the
sheet is tuned to work as a perfect absorber from one side, the
chirality parameter is zero and there is no possibility to tune the
reflection properties from the opposite side introducing chirality.

On the other hand, the omega coupling coefficient
$\amerh$ is not fixed by the total absorption
condition on one side, because from the first and third equations in
(\ref{eq:teem5}) we find \e \aeeoh={S\over
j\omega \eta_0}\mp {1\over \eta_0}\amerh
\l{om1}\f \e \ammoh={\eta_0 S\over j\omega }\pm
\eta_0\amerh.\l{om2} \f Comparing with
\r{both_sides}, we see that introducing omega coupling, we can
maintain the property of total absorption from one of the sides with
relaxed requirements on the electric and magnetic polarizabilities.
For instance, we can possibly engineer the omega coupling parameter
$\amerh$ so that the required magnetic
polarizability is much smaller than that dictated by \r{both_sides}.
The reflection coefficient from the side opposite to the matched one
we find from (\ref{eq:teem6}): \e \mathbf{E}_{\rm r}=\pm
\frac{2j\omega}{S}
\aemrh\It\cdot\mathbf{E}_{\rm
inc}. \l{omega_rotation}\f Thus, varying the omega coupling
parameter, we can control the co-polarized reflection from the
opposite side of the sheet, maintaining the matching and total
absorption properties from one side.

More functionalities become available if we allow nonreciprocal
response of the particles. For simplicity, let us concentrate here
on the cases where the magnetoelectric coupling is only due to
nonreciprocity, assuming that the chirality and omega coupling
coefficients are zero (the effects of chirality and omega coupling
have been considered above). In these cases, the coupling
coefficients satisfy \e
\aemoh=\ameoh,\qquad
\aemrh=-\amerh,\f
corresponding to Tellegen and ``moving'' particles, respectively
\cite{classes,basic}. From the first equation in (\ref{eq:teem5}),
we find that $\eta_0\aeeoh=\frac{1}{\eta_0}
\ammoh$ (Huygens' relation). From this and the
third relation we get \e \aeeoh=  {S\over
j\omega \eta_0}\mp {1\over
\eta_0}\aemrh.\l{mov_co_cr}\f The second and the
last relations in (\ref{eq:teem5}) connect the anti-symmetric parts
of the electric and magnetic polarizabilities with the Tellegen
parameter: \e \aeerh=\pm {1\over \eta_0}
\aemoh,\qquad \ammrh=\mp
\eta_0 \aemoh.\f Thus, if the Tellegen coupling
is present, its effects should be balanced with the nonreciprocity
in both electric and magnetic polarizabilities. Tellegen coupling
allows control of the reflection coefficient from the opposite side,
since \e \mathbf{E}_{\rm r}=\mp \frac{2j\omega}{S}
\aemoh\Jt\cdot\mathbf{E}_{\rm
inc}.\f We see that the Tellegen sheet can be designed to work as a
perfect absorber from one side and a twist polarizer in reflection
from the other side.

Finally, the antisymmetric part of the nonreciprocal coupling
coefficient allows to control the transmission coefficient from the
opposite side (for nonreciprocal sheets the transmission coefficient
is not anymore necessarily symmetric): \e \mathbf{E}_{\rm t}=\pm
\frac{2j\omega}{S}
\aemrh\It\cdot\mathbf{E}_{\rm
inc}.\f This transmission coefficient equals unity if
$\aemrh=\pm S/(2j\omega)$, in which case
equation \r{mov_co_cr} shows that all the polarizabilities are in
balance:
\begin{equation}
\eta_0\aeeoh=\pm\aemrh=\frac{1}{\eta_0}\ammoh=\frac{S}{j2\omega}.
\end{equation}
Using (\ref{eq:l1}), it is easy to show that the polarizabilities for
each individual particle should also be in balance  and equal to
\begin{equation}
\displaystyle\eta_0\aeeo=\pm\aemr=\frac{1}{\eta_0}\ammo={\eta_0\over 2}\displaystyle\frac{1}{\displaystyle
\frac{j\omega\eta_0}{S}+\beta_{\rm e}}.
\end{equation}

We see that the required electric and magnetic effective
polarizabilities are twice as small as compared to the simple case
of isotropic dipole particles \r{both_sides}. However, the resulting
amplitudes of the induced dipole moments and the  amplitudes of the
secondary plane waves are the same, because both moments are
generated by both incident fields. If this nonreciprocal array  is
excited from the absorbing side, these secondary plane waves cancel
the incident wave behind the sheet and they cancel each other in
the reflection direction, same as for the simple isotropic array.
But for the excitation from the opposite side, the induced dipole
moments are zero, because contributions due to the applied electric
and magnetic fields cancel out, and the sheet is transparent. We can
conclude that this interesting structure has the property of the
ultimately thin (a single layer of dipole particles) isolator: from
one side it acts as a total absorber while from the other side, the
sheet is transparent. Moreover, it appears that this is the only
possible configuration having this property.

\section{Uniaxial bi-anisotropic particles as components of totally absorbing arrays}

Next we will discuss some possible designs of bi-anisotropic
particles with the properties required for single-layer perfect
absorbers. From the reciprocal classes, the most interesting and
practically useful property is the omega coupling, since this effect
gives flexibility in the requirements on the electric and magnetic
polarizabilities and allows control over the reflection coefficient
from the back side of the absorbing sheet (see
\r{om1}--\r{omega_rotation}).

\subsection{Wire omega particles}

The classical topology of bi-anisotropic particles with  omega
coupling is an $\Omega$-shaped particle \cite{Saad,proposed,basic}.
It is clear that for a single uniaxial omega particle made of
a conducting wire (see picture in Table~\ref{ta:load_values}), in
approximation of electrically small particles, polarizabilities are
such that \cite{basic},
\begin{equation}
\displaystyle
\aeeo\ammo=-\aemr\amer=-(\aemr)^2.
\label{eq:om4}
\vspace*{.2cm}\\\displaystyle
\end{equation}
\begin{table*}[!t]
\centering
\caption{Conditions for perfect absorption}
\begin{tabular}{|p{50mm}|p{50mm}|p{50mm}|}
\hline
\rowcolor[gray]{.9}
\multicolumn{3}{|c|}{{\bf Condition for total absorption}}
\\
\hline
\rowcolor[gray]{.9}
Wire Omega
&
Omega---Tellegen
&
Chiral---Moving
 \\
\hline

\vspace{0.5mm} \includegraphics[width=0.3\textwidth]{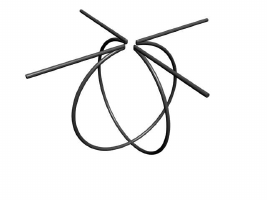}
\hspace*{1.4cm}
$
\displaystyle
\aeeoh=-\frac{1}{\eta_0^2}\ammoh$

&
\vspace{0.5mm} \includegraphics[width=0.3\textwidth]{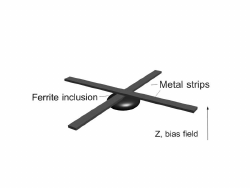}
\hspace*{.1cm}
$
\begin{array}{c}
\displaystyle
\eta_0\aeeoh\pm 2\aemrh-\frac{1}{\eta_0} \ammoh=0
\vspace*{.2cm}\\\displaystyle
\eta_0\aeerh\mp 2\aemoh-\frac{1}{\eta_0} \ammrh=0
\vspace*{.2cm}\\\displaystyle
\eta_0\aeeoh+\frac{1}{\eta_0}\ammoh=\frac{2S}{j\omega}
\vspace*{.2cm}\\\displaystyle
\eta_0\aeerh+\frac{1}{\eta_0} \ammrh=0
\end{array}$

& \vspace{0.5mm} \includegraphics[width=0.3\textwidth]{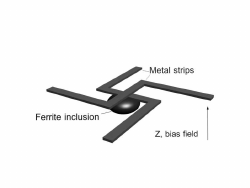}

 $
\begin{array}{c}
\displaystyle
\eta_0\aeeoh-\frac{1}{\eta_0} \ammoh=0
\vspace*{.2cm}\\\displaystyle
\eta_0\aeerh-\frac{1}{\eta_0} \ammrh=0
\vspace*{.2cm}\\\displaystyle
\eta_0\aeeoh\pm2\aemrh+\frac{1}{\eta_0}\ammoh=\frac{2S}{j\omega}
\vspace*{.2cm}\\\displaystyle
\eta_0\aeerh\mp2\aemoh+\frac{1}{\eta_0} \ammrh=0\vspace*{.5cm}
\end{array}$
\\
\hline \multicolumn{3}{|c|}{{\bf Reflected and transmitted fields
from the other side of a single-sided perfect absorber}}

\\  \hline
\rowcolor[gray]{.9}
Omega
&
Tellegen
&
Moving
\\
\hline

\vspace*{.1cm}\hspace*{.6cm}
$
\begin{array}{c}
\displaystyle\hspace*{-.2cm}
\mathbf{E}_{\rm r}=\pm
\frac{2j\omega}{S}
\aemrh\It\cdot\mathbf{E}_{\rm
inc}
\vspace*{.5cm}\\\displaystyle\hspace*{-.2cm}
\mathbf{E}_{\rm t}=0\vspace*{.5cm}
\end{array}$

&

\vspace*{.1cm}\hspace*{.6cm}
$ \begin{array}{c}

\displaystyle\hspace*{-.2cm}
\mathbf{E}_{\rm r}=\mp \frac{2j\omega}{S}
\aemoh\Jt\cdot\mathbf{E}_{\rm
inc}
\vspace*{.5cm}\\\displaystyle\hspace*{-.2cm}
\mathbf{E}_{\rm t}=0\vspace*{.5cm}
\end{array} $

&

\vspace*{.1cm}\hspace*{.6cm}
$ \begin{array}{c}
\displaystyle\hspace*{-.2cm}
\mathbf{E}_{\rm r}=0\vspace*{.5cm}\\\displaystyle
\mathbf{E}_{\rm t}=\pm
\frac{2j\omega}{S}
\aemrh\It\cdot\mathbf{E}_{\rm
inc}\vspace*{.5cm}
\end{array} $
\\
\hline

\end{tabular}
\label{ta:load_values}
\end{table*}
This condition is  a limitation on electromagnetic properties of a
wire omega particle. Using (\ref{eq:l1}), we find that the effective
polarizabilities of omega particles forming a periodical array
satisfy the same relation
\begin{equation}
\displaystyle
\aeeoh\ammoh=-(\aemrh)^2.
\label{eq:om5}
\vspace*{.2cm}\\\displaystyle
\end{equation}
Let us consider the limitation  (\ref{eq:om5}) together with the
first condition for total absorption in (\ref{eq:teem5}) (which is
the condition for zero reflection from an array of omega particles).
Combining these two equations, we get
\begin{equation}
\displaystyle
\aeeoh\pm \frac{2 j}{\eta_0}\sqrt{\aeeoh \ammoh}-\frac{1}{\eta_0^2}\ammoh=0.
\label{eq:om6}
\vspace*{.2cm}\\\displaystyle
\end{equation}
From this simple quadratic equation one can obtain the relation
\begin{equation}
\displaystyle
\aeeoh=-\frac{1}{\eta_0^2}\ammoh
\label{eq:om7}
\end{equation}
and using (\ref{eq:l1}), we get the same relation between the
polarizabilities of individual particles in free space
$\left(\aeeo=-\frac{1}{\eta_0^2}\ammo\right)$,
but this equation does not hold for passive omega particles, because
the different signs of the imaginary parts of the electric and
magnetic polarizabilities mean that the particle should be active.
On the other hand, it is impossible to satisfy the third condition
from (\ref{eq:teem5}) taking the limitation of (\ref{eq:om7}) into
account. Therefore, a wire omega particles cannot be used for the
design of perfect absorbers of this type. This is an interesting
fact because earlier nearly-total absorption was predicted in
structures which behave like omega particles \cite{mohammad}.
However, there is a significant difference between the case studied
in \cite{mohammad} and wire omega particles. For a wire omega
particle all the polarizabilities have the same resonance frequency,
but for the structure in \cite{mohammad}, one can tune the
structural parameters so that different polarizabilities  have
different resonance frequencies. Thus, it appears possible to break
the  limitation of (\ref{eq:om7}) using other kinds of omega
particles and achieve total absorption with the help of the
omega-coupling phenomenon.

\subsection{Omega-Tellegen particles}

Within the nonreciprocal classes, the most interesting properties
are the possibilities offered by nonreciprocal field coupling
phenomena in array of particles. Realization of such particles requires
inclusions of some nonreciprocal elements. The known structures for
the microwave frequency range \cite{basic} include magnetized
ferrite spheres coupled to specially shaped metal elements, see
illustrations in Table~1. However, both these structures exhibit
also reciprocal field coupling effects in addition to the desired
nonreciprocal effects.

A single uniaxial Tellegen particle shows also some omega field
coupling due to the asymmetrical position of the metal strips with
respect to the center of the ferrite sphere. For this reason, we call
it omega-Tellegen particle. Its polarizability dyadics have the form
\begin{equation}
\left\{\begin{array}{l}
\displaystyle
\aeebh=\aeeoh\It+\aeerh\Jt
\vspace*{.2cm}\\\displaystyle
\ammbh=\ammoh\It+\ammrh\Jt
\vspace*{.2cm}\\\displaystyle
\aembh=\amebh=\aemoh\It+\aemrh\Jt.
\end{array}\right. \label{eq:te3}
\end{equation}
Using relations (\ref{eq:u1}) and (\ref{eq:te3}), we get the
following conditions for total absorption in omega-Tellegen arrays
\begin{equation}
\begin{array}{c}
\displaystyle
\eta_0\aeeoh\pm 2\aemrh-\frac{1}{\eta_0} \ammoh=0
\vspace*{.2cm}\\\displaystyle
\eta_0\aeerh\mp 2\aemoh-\frac{1}{\eta_0} \ammrh=0
\vspace*{.2cm}\\\displaystyle
\eta_0\aeeoh+\frac{1}{\eta_0}\ammoh=\frac{2S}{j\omega}
\vspace*{.2cm}\\\displaystyle
\eta_0\aeerh+\frac{1}{\eta_0} \ammrh=0.
\end{array}\label{eq:te4}
\end{equation}
This shows that if we want to use the advantages offered by Tellegen
coupling, we need to design the particle so that the omega coupling
coefficient $\aemrh$ is properly balanced with the
electric and magnetic polarizabilities.

\subsection{Chiral-moving particles}

Likewise, the known artificial moving particle
\cite{basic,mov1,mov2} (picture in Table~1) exhibits reciprocal
magnetoelectric coupling because of its chiral shape. The properties
of such particle can be modeled by the polarizability dyadics of the
form
\begin{equation}
\left\{\begin{array}{l}
\displaystyle
\aeebh=\aeeoh\It+\aeerh\Jt
\vspace*{.2cm}\\\displaystyle
\ammbh=\ammoh\It+\ammrh\Jt
\vspace*{.2cm}\\\displaystyle
\aembh=-\amebh=\aemoh\It+\aemrh\Jt.
\end{array}\right. \label{eq:mo3}
\end{equation}
Using the relations (\ref{eq:u1}) and (\ref{eq:mo3}), the conditions for total absorption in the chiral-moving slab read
\begin{equation}
\begin{array}{c}
\displaystyle
\eta_0\aeeoh-\frac{1}{\eta_0} \ammoh=0
\vspace*{.2cm}\\\displaystyle
\eta_0\aeerh-\frac{1}{\eta_0} \ammrh=0
\vspace*{.2cm}\\\displaystyle
\eta_0\aeeoh\pm2\aemrh+\frac{1}{\eta_0}\ammoh=\frac{2S}{j\omega}
\vspace*{.2cm}\\\displaystyle
\eta_0\aeerh\mp2\aemoh+\frac{1}{\eta_0} \ammrh=0.
\end{array}\label{eq:mo4}
\end{equation}
In this case, the chirality parameter $\aemoh$ should be
balanced with the anti-symmetric parts of the electric and magnetic
polarizabilities. Implementation  of total-absorption arrays using
omega-Tellegen or chiral-moving particles  presents significant
difficulties. To the best of our knowledge, there are no analytical
models to calculate the individual polarizabilities of these
particles. The relations between the effective and individual
polarizabilities for these particles become more involved. Finally,
the design presupposes the use of ferrites and magnetic field bias
which presents practical difficulties. On the other hand, these
topologies offer unique properties such as a thin sheet operating as
an isolator, and they clearly deserve further studies.

\section{Conclusions}

We have considered possible approaches for realization of perfect
absorbers using ultimately thin (single layers of particles)
structures. The thickness cannot be strictly zero (in the
electromagnetic sense) because we must allow magnetic response in
the layer. We have demonstrated that to realize total absorption
from both sides of the sheet, we just need to realize balanced
electric and magnetic polarizabilities and all magnetoelectric
polarizabilities should be zero. Further, we have considered
single-layer sheets which operate as perfect absorbers only when
illuminated from one of the two sides of the sheet and studied what
functionalities can be engineered for illumination from the opposite
side. We have shown that introducing  omega coupling in the
constituent particles makes it possible to realize a layer which
acts as a perfect absorber from one side with controllable
co-polarized reflection from the opposite side of the sheet. For
reciprocal structures, it has been shown that tuning the layer to
act as a perfect absorber from one side does not allow to have any
chirality in the layer. We have seen that allowing nonreciprocity in
the properties of the absorbing particles offers possibilities for
more functionalities. A Tellegen sheet can be designed to work as a
perfect absorber from one side and a twist polarizer in reflection
from the other side. The antisymmetric part of the nonreciprocal
coupling coefficient (i.e., a layer of particles with the
constitutive parameters of an artificial moving medium) makes it
possible to achieve total absorption from one side, but controlled
transmission coefficient from the opposite side. In particular, the
regime when the layer acts as a perfect absorber from one side,
while from the other side the sheet is transparent, is possible.
This corresponds to the ultimately thin isolator. Finally, we have
studied some particular examples of possible realizations of
single-layer perfect absorbers with the use of omega,
omega-Tellegen, and chiral-moving particles as canonical examples of
uniaxial bi-anisotropic particles.

The most interesting properties are offered by nonreciprocal
bi-anisotropic particles. They can be realized in practice as
near-field coupled magnetized ferrite inclusions  and metal strips
of wires, as was proposed in \cite{classes} (see pictures in
Table~\ref{ta:load_values}). Now we are developing analytical models
of omega-Tellegen and chiral moving particles which are expected to
enable design and optimization of inclusions for proposed
nonreciprocal absorbing layers.

\end{document}